\documentclass[letterpaper,11pt,onecolumn]{IEEEtran} 
\usepackage{amsmath,amsfonts}
\usepackage[utf8]{inputenc}
\usepackage{algorithmic}
\usepackage{algorithm}
\usepackage{array}
\usepackage[caption=false,font=normalsize,labelfont=sf,textfont=sf]{subfig}
\usepackage{textcomp}
\usepackage{stfloats}
\usepackage{url}
\usepackage{verbatim}
\usepackage{graphicx}
\usepackage{bpchem}
\begin{document}
\title{  Dielectric Modulated Double Gate Bilayer Electrode Organic Thin Film Transistor based Biosensor for Label-Free  Detection:Simulation Study and Sensitivity Analysis}
\author{Sushil Kumar Jain, \IEEEmembership{Member, IEEE}, and Amit Mahesh Joshi, \IEEEmembership{Member, IEEE}
}

\maketitle

\begin{abstract}
A dielectric-modulated double gate bilayer electrodes organic thin-film transistor (DMDGBE-OTFT) based sensor is proposed for label-free biomolecule detection. The double gate of DMDGBE-OTFT is used for creating two symmetrical gates underlap regions on both sides of the organic semiconductor. The parallel immobilization of biomolecule in two gates underlaps region changes the on-current ($I_{ON}$) of the DMDGBE-OTFT. Bilayer electrode is also used for significant reduction of barrier height to enhance the performance of the proposed device. The change in the drain current has been utilized to evaluate the sensitivity of the DMDGBE-OTFT for biomolecule recognition having different dielectric constants and corresponding charge densities using 2-D physics based numerical simulation when biomolecules are immobilized in the gate underlap area. The ATLAS TCAD tool is used to investigate the sensitivity performance of the DMDGBE-OTFT. The proposed DMDGBE-OTFT has 24.2\% higher sensitivity in comparison to the recently reported OTFT-based biosensors for label-free detection of biomolecules. The DMDGBE-OTFT biosensor has a lot of potential for flexible biosensing applications in the future because of its flexibility , high sensitivity,biocompatibility and  low cost.
\end{abstract}


\section{Introduction}
\IEEEPARstart{O}{rganic} bioelectronics has successfully bridged the gap between semiconductors and biological systems, producing low-cost, flexible, disposable, and lightweight organic biosensors \cite{Sun2021-ap}. Organic thin film transisitors (OTFTs) have been considered as one of the potential solutions for bio-sensing applications\cite{jain2020performance}. 
Traditional biosensing detection methods include blood tests, tumor markers, immunological assays, and genetic testing. 
There have been substantial growth of field-effect transistor (FET) based biomedical applications \cite{ahmad2022novel}. Among the several biosensor possibilities, biosensing with FETs has been identified as a promising choice. FETs were also recently utilized to diagnose the highly contagious COVID-19 illness due to the corona virus \cite{Seo2020-08-ir}. Various research groups \cite{Saha2021-ep,Dwivedi2020-ia,Hafiz1905-sg,Mahalaxmi2020-12-01-kt,Dixit2021-yd,Wadhwa2019-bg,Goel2021-lj,Priyadarshani2021-tj,Wang2019-bj,Narang2015-ho,Kanungo2017-gd,Kumar2020-up} have reported on the development of dielctrically modulated FET-based biosensors. Biosensors produced using FETs may be fabricated using either inorganic or organic semiconducting materials \cite{Seo2020-vl,Zaki_undated-dx}. Inorganic FET-based biosensors have demonstrated several advantages, including CMOS compatibility and scaling, as well as low-cost label-free detection. However, they have high-temperature processing and larger fabrication cost \cite{Singh2020-04-up}. Organic thin-film transistors (OTFT) are another type of  fabricated biosensor  using organic semiconducting material. Numerous organic semiconductors like  pentacene, \IUPAC{dinaptho[2,3{-}b:2{'},3{'}{-}f]thieno[3,2{-}b]thiophene} (DNTT) and P3HT, etc with chemical compositions are similar to different biological molecules. They have been extensively employed in the development of high-quality OTFTs applications. For a low cost, OTFT based sensors can be printed in a variety of ways on paper or plastic substrates \cite{Rashid2021-08-15-fk}. Organic biosensors solve the drawbacks of inorganic biosensors and can be incorporated with electrical circuits in addition to the detection of the biomolecule. Organic transistor-based biosensors provide several advantages, including enhanced parametric response, sensitivity, and high selectivity. The immobilisation of GOx (glucose oxidase) on dielectric layer has resulted in the development of OTFTs-based glucose biosensors \cite{Bartic2003-ev}. Sensitivity of glucose concentration has been determined by monitoring variations in the drain current as a result of the biocatalytic reaction between glucose and GOx. Khan et al. \cite{Khan2011-iq} used covalent attachment methodology  to develop a label-free detection of anti-BSA utilising a pentacene-based OTFT biosensor. Label-free histidine-rich protein biosensors with great sensitivity have been produced by using OTFTs \cite{Minamiki2017-iw}. It has been also successfully demonstrated that  pH sensors using OTFT can be built with pentacene semiconducing organic layers \cite{Loi2005-pk}. Label free DNA biosensor using  P3HT polymer is also reported  in which dsDNA and ssDNA are distinguished by employing gold electrodes to distinguish between the two types of DNA \cite{Yan2009-oc}. Label-Free Dielectrically modulated Metal trench OTFT as a biosensor was studied with improved sensitivity. Dielectrically modulated TFT-based biosensors have been proven to be very effective in detecting a variety of biomolecules in many recent studies \cite{Rashid2021-08-15-fk}. Apart from that, the dielectrically modulated FETs provide many benefits, including, the capacity to immobilize on a range of surfaces, low-cost label-free detection, and the potential to enable rapid identification with excellent affinity associating with the biomolecules \cite{joshi2022label}. However, it is still an challenges to have good sensitivity  using OTFTs in biosensor applications. The literature of various biosensors based on dielectric modulated field effect transistors (FETs) have been observed for detecting urea, lactate, glucose, biotin, DNA, streptavidin, ammonia, penicillin, and other biomolecules \cite{Goel2021-lj,Dixit2021-yd,Saha2021-ep,Syu2018-06-gd,Zan2012-zf}. In comparison to inorganic-based Dielectric modulated TFT, there has been relatively less simulation and modeling-related work done to evaluate the sensitivity of dielectrically modulated OTFT based biosensors. Additionally, FET-based biosensing is important in the medical, beverage, and food sectors, where low-cost, biocompatible, and adaptable biomolecule determination is highly demanding. We increased the drain current sensitivity of biosensors by design and modeling a novel OTFT structure using organic material (DNTT) \cite{sadighbayan2021laser,ahmad2021design,gupta2020surface}. 
The proposed device is called dielectric-modulated double gate bilayer electrodes organic thin-film transistor (DMDGBE-OTFT) because of double gate and bilayer of metal electrode for source and drain. Suggested devices improve many metrics, including drain on-current, conduction current density, and transconductance,  most crucially, sensitivity. 
The rest of the paper is organized as follows: The OTFT structure used in the device structure and simulation theory is explained in section II.  Section III presents the results and discussion. Section IV presents the conclusions taken from the investigation{'}s findings.

\begin{figure}[htbp]
\centerline{\includegraphics[width=0.7\columnwidth]{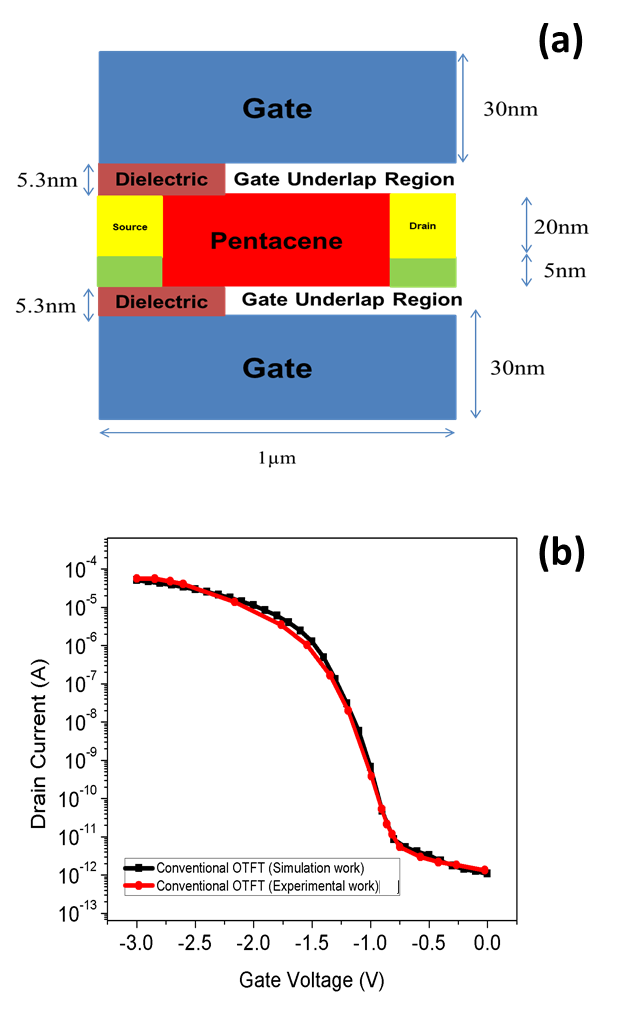}}
\caption{Structure of proposed DMDGBE-OTFT (b) Experimental transfer characteristic of conventional OTFT compared with the simulation characteristic [26].}
\label{fig1}
\end{figure}
\begin{figure}[htbp]
\centerline{\includegraphics[width=0.7\columnwidth]{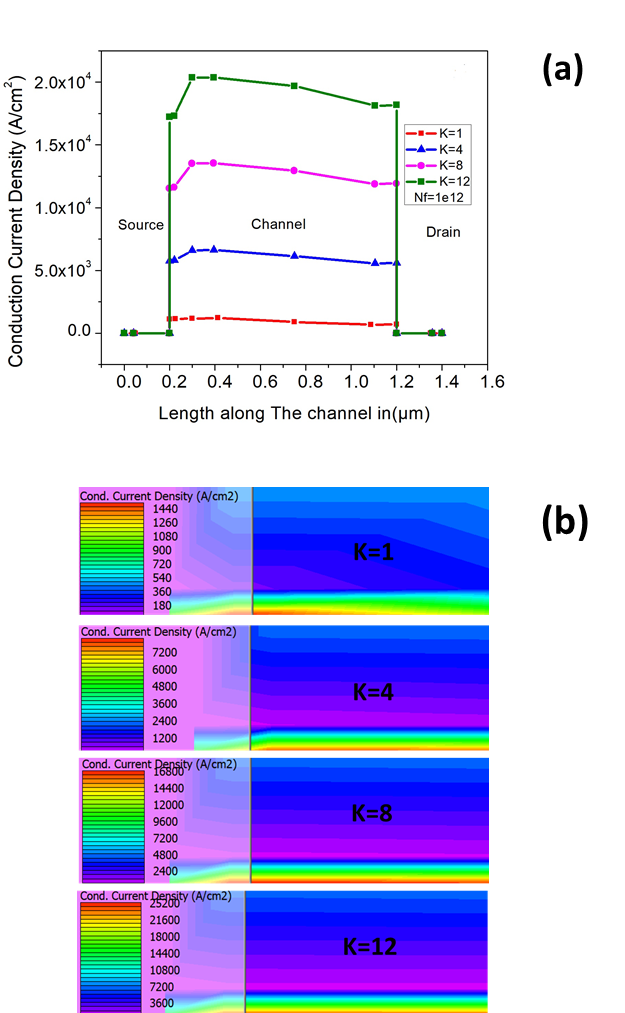}}
\caption{(a) Conduction current density for different K (b) Conduction current density profile along the channel length.}
\label{fig2}
\end{figure}

\begin{table}
\centering
\caption{DIMENSIONAL PARAMETERS  OF PROPOSED DEVICE}
\begin{tabular}{ l l }
    \hline
    \bf Parameter & \bf Value\\
    \hline
    Organic semiconductor (DNTT) thickness & 25nm \\
    Gate underlap Thickness ($T_g$ ) & 5.3nm to 2.3nm            \\
   Gate thickness & 30nm\\
   Source/Drain contact thickness & 25nm\\
   Gate underlap length ($L_g$) & 800 nm \\
   Dielectric thickness & 5.3nm \\
   
    \hline
\end{tabular}
\end{table}

\begin{table}[h!]
\centering
\caption{MATERIAL PARAMETERS OF PROPOSED DEVICE}
\begin{tabular}{ l l }
    \hline
    \bf Parameter & \bf Value\\
    \hline
   Relative permittivity of DNTT & 3.0 \\
    Affinity of DNTT & 1.81 eV          \\
   Band gap of DNTT & 3.38 eV\\
   P-type doping  & $1 \times 10^{16}cm^{-3}$\\
  Work function of gold & 5.0 eV\\
   Work function of Platinum  & 5.7 eV \\
   Dielectric constant & 3.37 eV \\
   
    \hline
\end{tabular}
\end{table}

\begin{table}
\centering
\caption{DIELECTRIC CONSTANT OF BIOMOLECULES}
\begin{tabular}{ l l l}
    \hline
    \bf Biomolecule & \bf Dielectric Constant  & \bf References\\
    \hline
   Protein & 2.50 & \cite{Wadhwa2018-05-uk} \\
   Biotin &	2.63 & \cite{Wadhwa2018-05-uk}        \\
   uricase  & 	1.54  & \cite{Mahalaxmi2020-12-01-kt} \\
  Bacteriophage  & 	6.3	 & \cite{Mahalaxmi2020-12-01-kt} \\
  Gelatin  & 	12  & \cite{Mahalaxmi2020-12-01-kt} \\

       \hline
\end{tabular}
\label{Tab11}
\end{table}
\begin{table}[h!]
\centering
\caption{COMPARISON OF OUR WORK WITH PREVIOUS REPORTED WORK}
\begin{tabular}{ l l l l }
    \hline
    \bf Ref. & \bf V$_{DS}$/V$_{GS}$(V)  & \bf Permittivity & \bf Sensitivity \\
    \hline
  \cite{Hafiz2019-02-dd}	& 1.0/1.0 & 	10	 & 1.2 \\
   
  \cite{Kumar2020-up} &	0.5/1.0 &	8 &	1.385   \\
   
  \cite{Venkatesh2017-09-vs} & 0.5/1.0 &    12 & 7.3 \\
   
  \cite{Singh2017-lg} & 0.5/1.5 &    10 & 6 \\
   
  \cite{Rashid2021-08-15-fk} & -1.5/-3.0 &    12 & 12.57 \\
   
    [This work] & -1.5/-3.0 &    12 & 15.61 \\
   
      \hline
\end{tabular}
\end{table}

\section{Device Structure}
Fig. 1(a) shows the proposed novel \IUPAC{dinaptho[2,3{-}b:2{'},3{'}{-}f]thieno[3,2{-}b]thiophene}  (DNTT) based DMDGBE-OTFT device. The DMDGBE-OTFT uses double gates and bi-layer metal contacts to improve sensitivity and performance over a conventional OTFT based sensor. The AlOx/SAM (K=3.37) gate dielectric \cite{Kraft2016-we} is ultra-thin (5.3nm) and intended for high capacitance and low voltage. Many authors \cite{Im2007-ha,Wadhwa2018-yy} have reported that the size of the biomolecules (protein,biotin,gelatin etc.) lies generally within 5 nm cavity thickness. Table I lists the structural specifications of proposed biosensor. Table II outlines the materials properties used in DMDGBE-OTFT simulation. The organic semiconducting material (DNTT) has 3, 1.81eV, and 3.38eV relative permittivity, hole affinity, and energy band gap, respectively \cite{Zaki_undated-dx}. The two gate underlap region is created in the proposed device to immobilise biomolecules, $T_g$ is the thickness of the gate underlap area, where biomolecules are immobilised. SILVACO TCAD has evaluated the p-type gate underlaps DMDGBE-OTFT. Density of defect states (DOS) and poole-frenkel models are used to simulate charge transfer carrier hopping in organic semiconductors like pentacene or DNTT. The parameters used in simulation were calibrated to match the experimental characteristics given in literature (as shown in Fig. 1(b)) \cite{Kraft2016-we}.

\begin{figure}[htbp]
\centerline{\includegraphics[width=0.6\columnwidth]{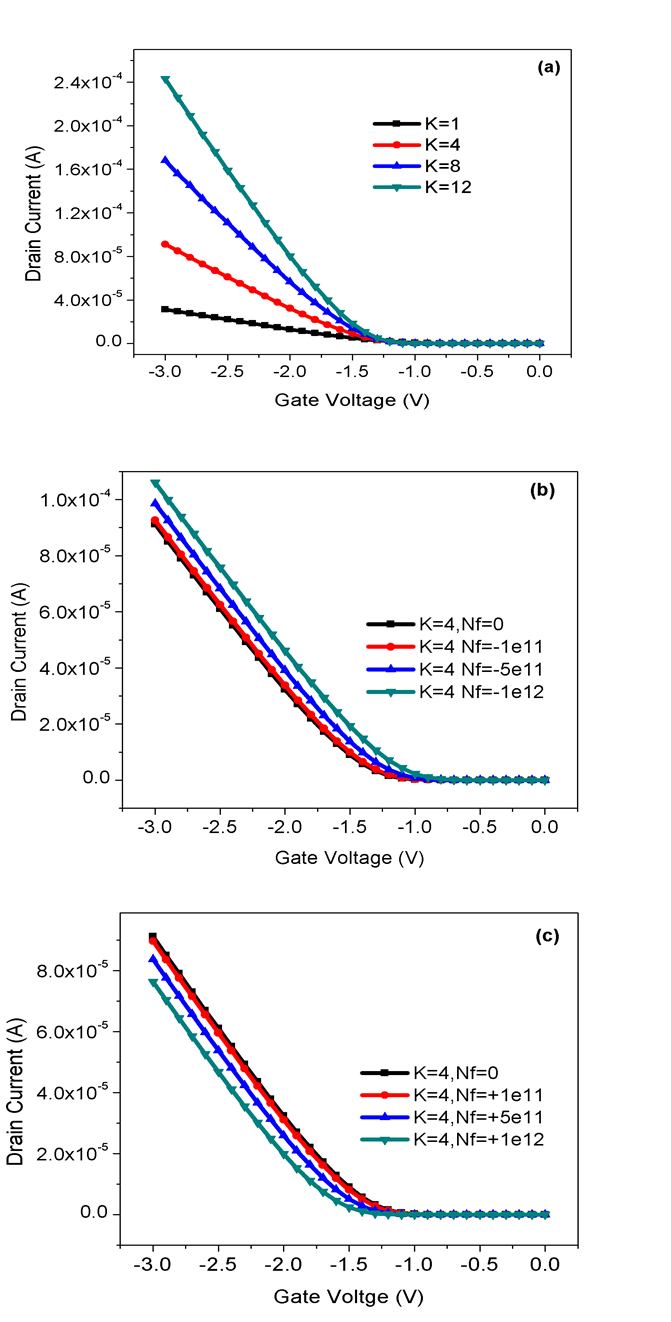}}
\caption{Transfer characteristics at V$_{DS}$= -1.5V (a) with $T_g$=5.3nm and varying dielectric constant (K = 1, 4, 8, 12). (b) with $T_g$=5.3nm and varying negative charge of the biomolecule. (c) with $T_g$=5.3nm and varying positive charge of the biomolecule.}
\label{fig3}
\end{figure}
\vspace{-0.5cm}
\begin{figure}[htbp]
\centerline{\includegraphics[width=0.6\columnwidth]{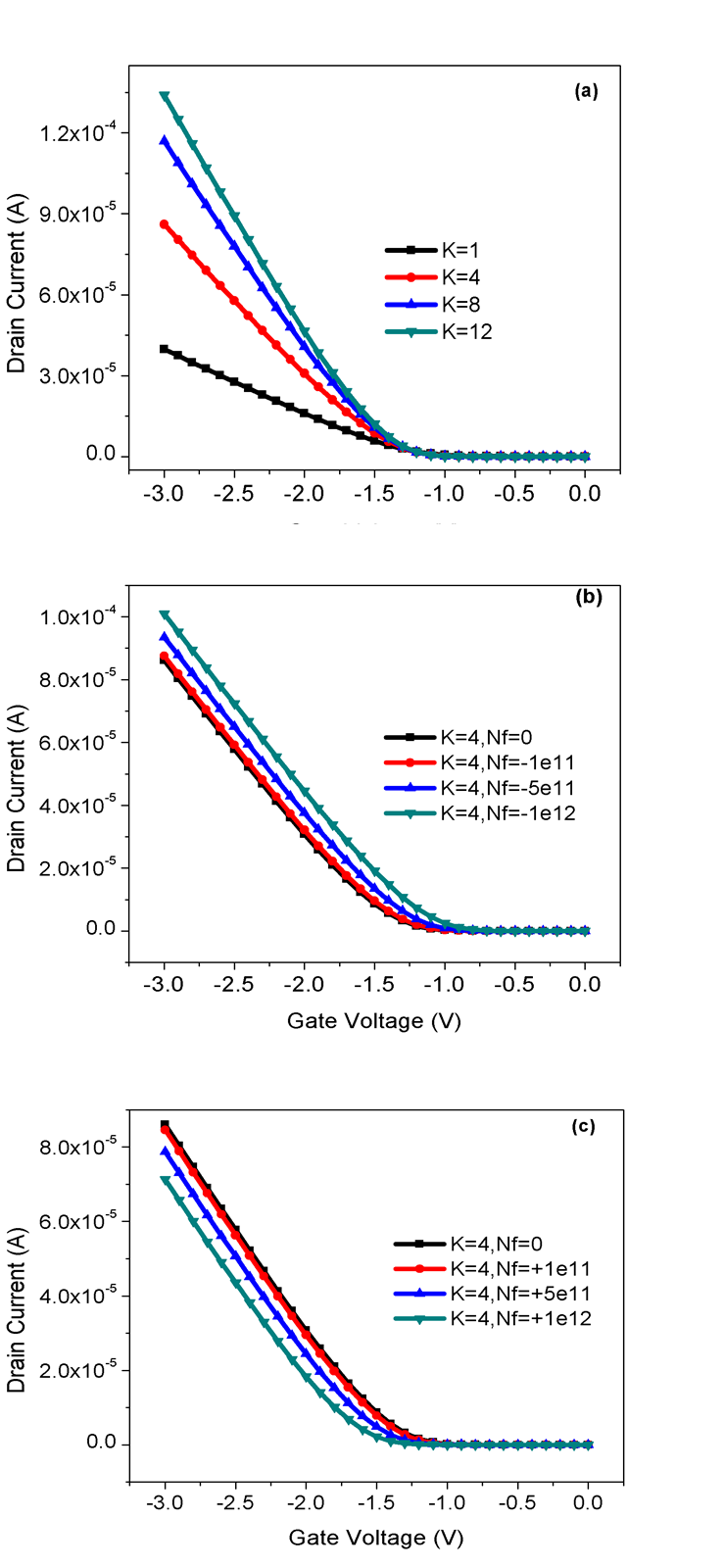}}
\caption{Transfer characteristics at V$_{DS}$= -1.5V (a) with T$_g$=3.3nm and varying dielectric constant (K = 1, 4, 8, 12). (b) With T$_g$=3.3nm and varying negative charge of the biomolecule. (c) With T$_g$=3.3nm and varying positive charge of the biomolecule.}
\label{fig4}
\vspace{-0.5cm}
\end{figure}
\begin{figure}[htbp]
\centerline{\includegraphics[width=0.55\columnwidth]{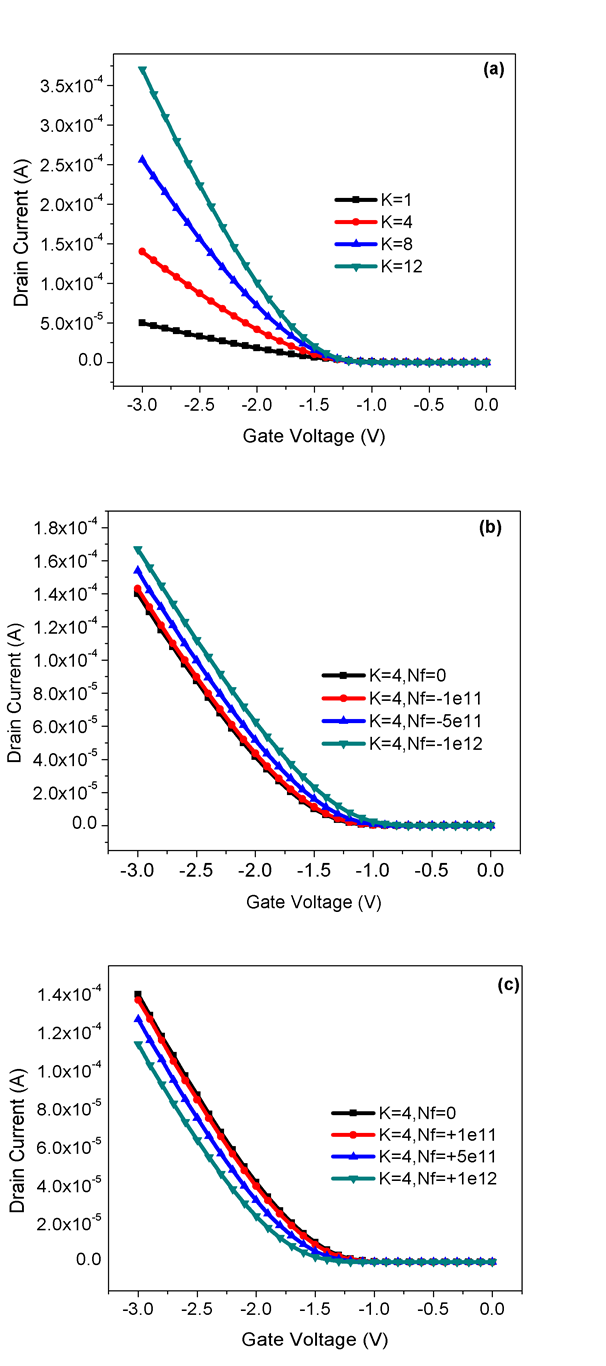}}
\caption{Transfer characteristics at V$_{DS}$= -3.0V (a) with T$_g$=5.3nm and varying dielectric constant (K = 1, 4, 8, 12). (b) With T$_g$=5.3nm and varying negative charge of the biomolecule. (c) With T$_g$=5.3nm and varying positive charge of the biomolecule.}
\label{fig5}
\vspace{-0.5cm}
\end{figure}
\begin{figure}[htbp]
\centerline{\includegraphics[width=0.6\columnwidth]{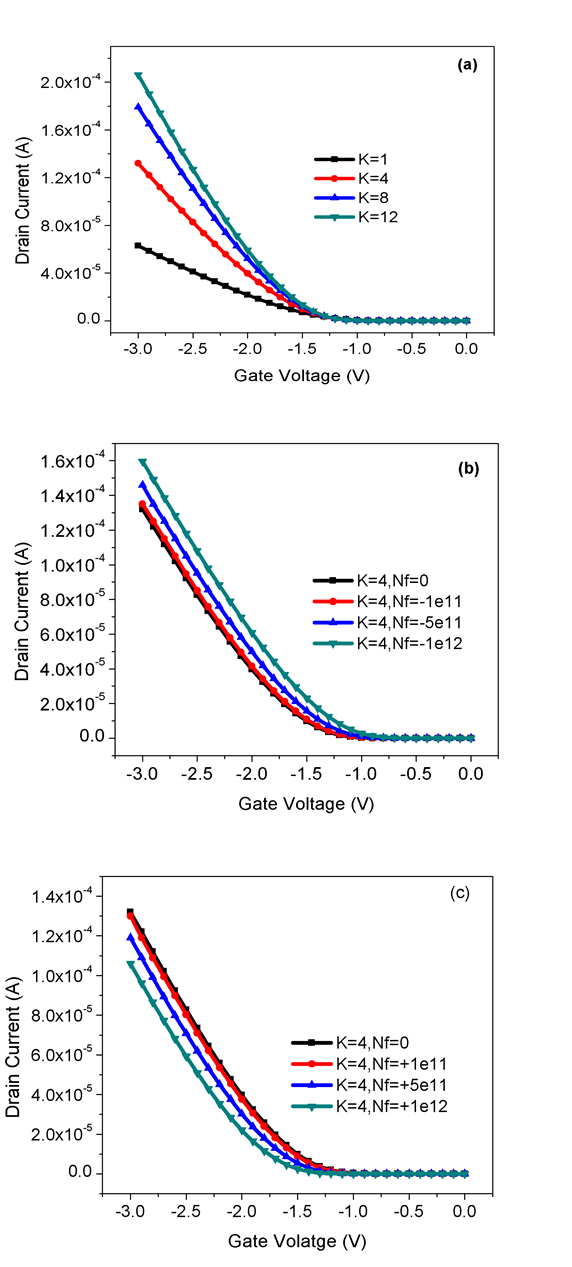}}
\caption{Transfer characteristics at V$_{DS}$= -3.0V (a) with T$_g$=3.3nm and varying dielectric constant (K = 1, 4, 8, 12) (b) with T$_g$=3.3nm and varying negative charge of biomolecule. (c) with T$_g$=3.3nm and varying positive charge of the biomolecule.}
\label{fig6}
\vspace{-0.5cm}
\end{figure}

\begin{figure}[htbp]
\centerline{\includegraphics[width=0.6\columnwidth]{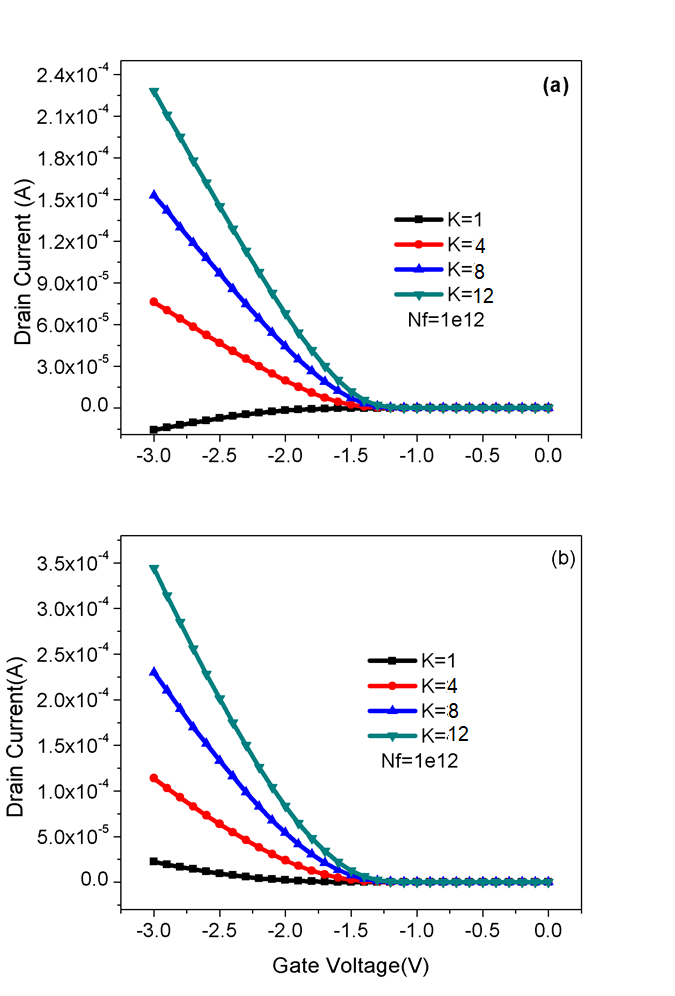}}
\caption{Transfer characteristics with charge density  Nf= $1x10^{12}cm^{-2}$ (a) by changing dielectric constant (K = 1, 4, 8, 12) at V$_{DS}$=-1.5V (b) By changing dielectric constant (K = 1, 4, 8, 12) at V$_{DS}$=-3.0V.}
\label{fig7}
\vspace{-0.5cm}
\end{figure}

\begin{figure}[htbp]
\centerline{\includegraphics[width=0.6\columnwidth]{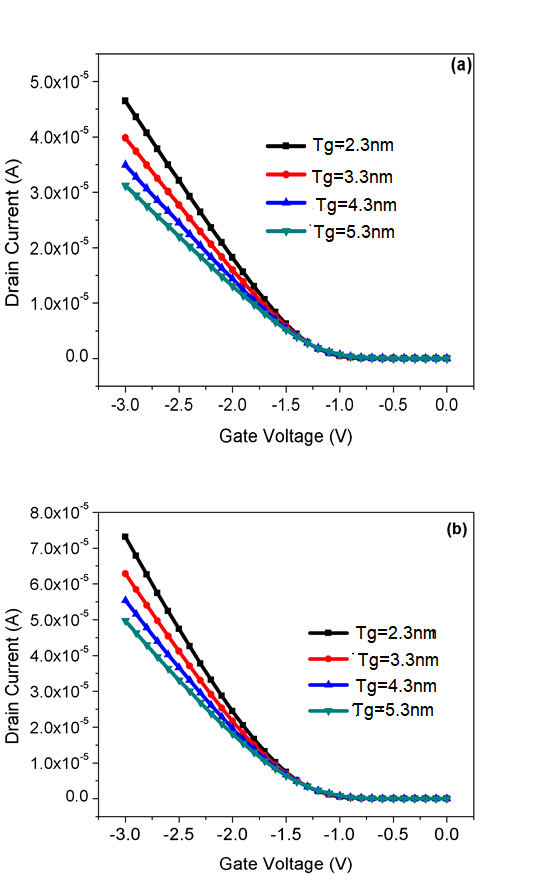}}
\caption{Effect of gate underlap thickness on drain current for K=4 (a) at V$_{DS}$ = -1.5 V and (b) at V$_{DS}$ = -3.0 V.}
\label{fig8}
\label{fig11}
\vspace{-0.5cm}
\end{figure}

\begin{figure}[htbp]
\centerline{\includegraphics[width=0.6\columnwidth]{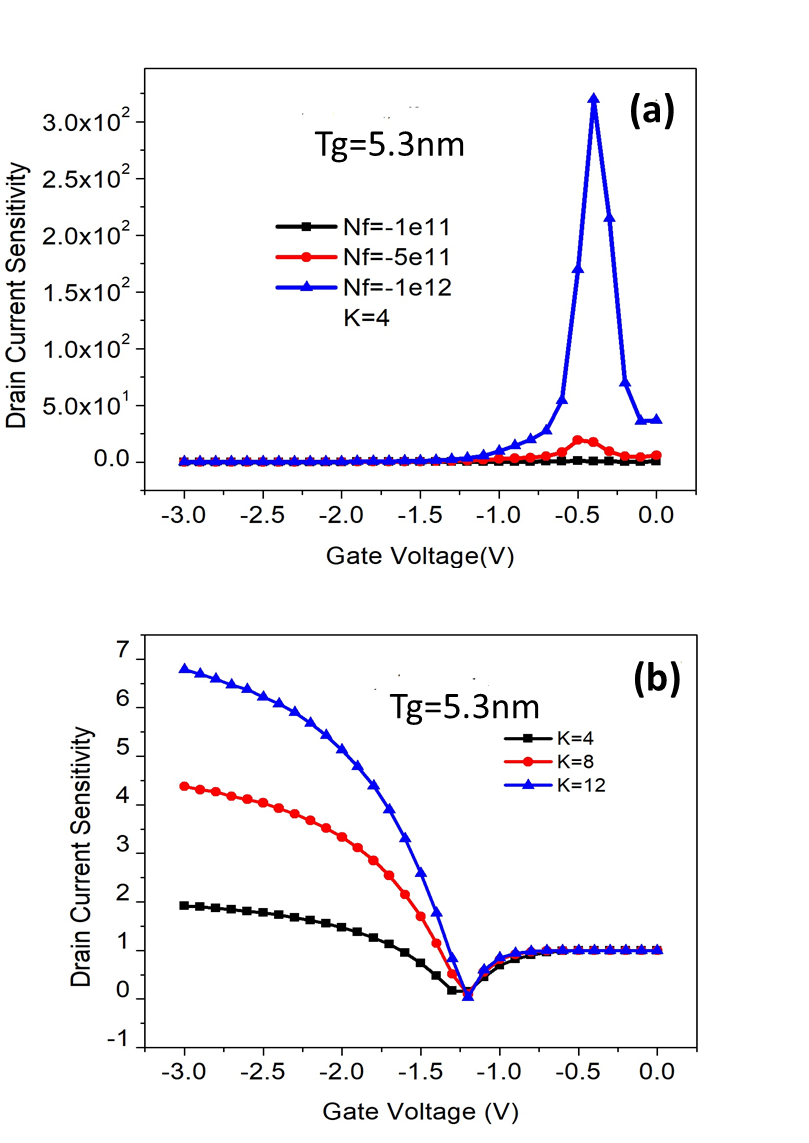}}
\caption{(a) Drain current sensitivity for K=4 at T$_g$=5.3nm for different negative charged molecule  at V$_{DS}$ = -1.5 V (b) Drain current sensitivity for different K at T$_g$=5.3nm at V$_{DS}$ = -1.5V.}
\label{fig9}
\label{fig11}
\vspace{-0.5cm}
\end{figure}

\section{Results and Discussions}
\label{Results and Discussions}
Biomolecules found in the environment can be classified as either charged or neutral. The effect of a neutral biomolecule on performance characteristics can be simulated using only its dielectric constant, whereas the effect of a charged biomolecule requires both its charge density and dielectric constant. For label-free electrical detection using biomolecules such as DNA, protein, uricase, biotin, and gelatin, the charge density ($N_f$) of biomolecules reported in literature varies from $-1 \times 10^{11}cm^{-2}$   to   $+1 \times 10^{12}cm^{-2}$ \cite{Singh2020-04-up}. The range of dielectric constants is considered in the range of K=1 to K=12 \cite{Wadhwa2018-05-uk}. Table III shows the dielectric constants of a few biomolecules.
\subsection{The Effect of Bimolecular Charge and Dielectric Constant of biomolecule on Conduction Current Density}
The  biomolecule becomes immobilized in the gate underlap area of the proposed device. When biomolecules with constant charge density $1 \times 10^{12}cm^{-2}$ and varying K (1 to 12) are attached in the cavity, the charge density in the channel area of the organic material varies. As the dielectric constants of bio-molecules are increased, the carrier concentration in the channel area increases, resulting in a higher conduction current density, as seen in Fig. 2. Finally, due to changes in the conduction current density, the drain current of the proposed device would change. Maximum conduction current density is attained at maximum K=12 because to the greater dielectric constant of biomolecule.

\subsection{ The Effect of Bimolecular Charge and Dielectric Constant of biomolecule on Drain Current}
The changes in drain current characteristics caused by the binding of various neutral biomolecules with varying dielectric constants in the gate underlap cavity region at $V_{GS}$=-3.0 and $V_{DS}$=-1.5V are depicted in Fig. 3(a), 4(a), 5(a), and 6(a). When the dielectric constant of the underlap area grows, the on-current ($I_{ON}$) increases because the conduction current density improves.
The drain current characteristics for varying underlap thicknesses ($T_g$) when negative-charged biomolecules are connected in the underlap area at $V_{GS}$=-3.0 and $V_{DS}$=-1.5V are shown in Fig. 3(b), 4(b), 5(b), and 6(b). The on-current ($I_{ON}$) grows as the negative charge density of biomolecules increases. Fig. 3(c), 4(c), 5(c), and 6(c) show that as the positive charge density increases, the DMDGBE-OTFT gate overlap has the opposite effect on the on-current ($I_{ON}$), causing the drain current to decrease. Because the biomolecules adsorbed in the cavity area have a positive (negative) charge, the barrier between the conduction band of the channel and the valence band of the source widens (shrinks). The drain current fluctuates dramatically when biomolecules have K=12 and maximum positive charge density, as shown in Fig. 6 (a) and 6 (b). With an underlap thickness of 5.3 nm and K = 12, the on-current ($I_{ON}$) may achieve its maximum value at $V_{GS}$=-3.0V, but it is lower at $V_{GS}$=-1.5V, as shown in Fig. 7.

\subsection{Variation in Device Performance Due to Changes in Gate Underlap Thickness ($T_g$) }
Fig. 8(a) and 8(b) show how drain current varies with gate underlap thickness ($T_g$) while other dimensional parameters remain constant. When the underlap thickness ($T_g$) grows from 2.3 nm to 5.3 nm, the drain current variation is reduced. The drain current achieves its maximum value when the underlap thickness ($T_g$) is 2.3 nm, as shown in Fig. 8. When $T_g$ is raised, drain current decreases, which might be due to a higher barrier at the source–channel boundary. We add a bilayer electrode into the device to reduce it as much as feasible. There is less variance in drain current when V$_{DS}$ =-1.5V is employed instead of V$_{DS}$ =-3.0V.
\begin{figure}[htbp]
\centerline{\includegraphics[width=0.6\columnwidth]{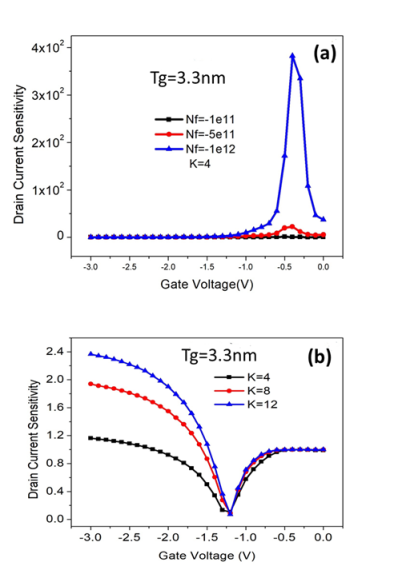}}
\caption{(a) Drain current sensitivity for K=4 at T$_g$=3.3nm for different negative charged molecule  at V$_{DS}$ = -1.5 V (b) Drain current sensitivity for different K at T$_g$=3.3nm at V$_{DS}$ = -1.5 V.}
\label{fig10}
\label{fig11}
\vspace{-0.5cm}
\end{figure}

\begin{figure}[htbp]
\centerline{\includegraphics[width=0.6\columnwidth]{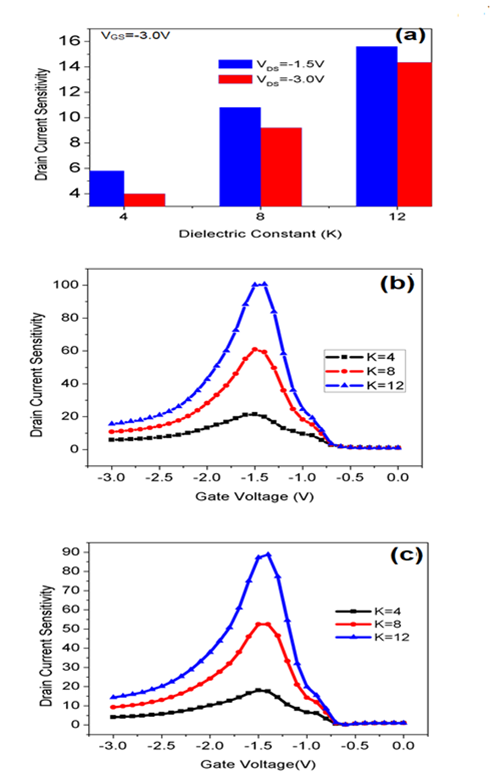}}
\caption{(a) comparison of Drain current sensitivity  with variation in dielectric constant for charged biomolecule(b)Drain current sensitivity variation with V$_{GS}$  at V$_{DS}$ = -1.5V (c) Drain current sensitivity variation with V$_{GS}$  at V$_{DS}$ = -3.0V}
\label{fig11}
\vspace{-0.5cm}
\end{figure}
\subsection{Drain Current sensitivity}


Drain current sensitivity is used to calculate the efficiency of biological sensors in the following manner: The sensitivity of the proposed device may be computed using the formula below when both neutral and charged biological molecules are attached in the gate underlap cavity region.
\begin{equation}
\centering
Sensitivity =   I_{ON}  (bio) - I_{ON} (air)/I_{ON} (air)
\end{equation}
When the gate underlap area is filled with biomolecules, $I_{ON}$(bio) equals drain current; when the gate underlap area is empty, $I_{ON}$ (air) equals drain current. Equation (1) demonstrates the need for a considerable difference between $I_{ON}$ (bio) and $I_{ON}$ (air)  to detect the target and improve sensitivity for the biosensing application. For modeling the absence of a biomolecule, researchers have calculated the sensitivity in the literature using an air environment (K=1). TABLE IV compares current biosensing research to previous reported results. As the dielectric constant of the gate underlap cavity region changes, so does the current density in the channel. The device sensitivity improves with the change in charge concentration of the channel. Fig. 9 and 10 depict the relationship between drain current sensitivity and gate voltage for biomolecules with various dielectric constants and charge densities at different thickness of gate underlap. The results illustrate that the suggested device{'}s has higher sensitivity. As demonstrated by the wide range of sensitivity, the proposed DMDGBE-OTFT is suitable for label-free biosensors to identify the bio-molecules for future flexible electronics.Fig. 9(a) and 10(a) depict the effect of variation in biomolecule negative charged density on drain current sensitivity with  dielectric constant K=4 at $T_g$=5.3nm and $T_g$=3.3nm, respectively. The peak sensitivity was obtained at V$_{GS}$ =-0.5V. Fig. 9(b) and 10(b) show the effect of several neutral biological molecules on drain current sensitivity when the gate underlap area ($T_g$) thickness is 5.3nm and 3.3nm, respectively. The sensitivity of the device improves as the relative permittivity increases after V$_{GS}$ =-1.2V for dielectric constants ranging from 1 to 12.
Fig.11 (a) depicts the variations in drain current sensitivity for charged biomolecules as dielectric constants vary from K=4 to K=12 with constant charge density ($N_f$=$+1 \times 10^{12}cm^{-2}$). Charged biomolecules with K=12 exhibit a comparatively high drain current sensitivity of 15.91 at $V_{GS}$=-3.0V and $V_{DS}$=-1.5V compared to earlier research \cite{Rashid2021-08-15-fk}. Fig. 11(b) and 11(c) demonstrate how gate to source biasing voltage affects drain current sensitivity for charged biomolecules with varying dielectric constants at $V_{GS}$=-1.5V and $V_{DS}$=-3.0V, respectively. The drain current sensitivity of charged biomolecules rises as the dielectric constant increases due to differences in current density.

\section{Conclusion}
The DMDGBE-OTFT is a label-free biosensing device that has been designed and simulated. In order to replicate the presence of a biomolecule in the gate underlap cavity area, the dielectric constant is varied from K= 1 to 12, which corresponds to various biomolecules. The magnitude and profile of conduction current density inside the device is varied for different dielectric constants. The effect of dielectric constant, gate underlap thickness, and biomolecule charge (positive and negative) on drain current has been investigated. As the dielectric constant of the neutral biomolecule increases, so does the on-current ($I_{ON}$). When the negative charge density of charged biomolecules with constant dielectric constant increases, the on-current ($I_{ON}$) rises as well. When the positive charge density increases, the DMDGBE-OTFT exhibits the opposite behavior for on-current ($I_{ON}$), resulting in a decrease in drain current. The simulated findings show that the gate underlap width grows, the current sensitivity drops at stable points $V_{GS}$=-3.0V and $V_{DS}$=-1.5V. The selection of double gate and bilayer electrodes is also important in obtaining a significantly improved value of sensitivity. The sensitivity of the proposed device is 24.22\% is higher in comparison to recently reported work. As a result, the proposed device has a lot of potential for future flexible and bio-compatible OTFT based biosensors.

\section*{Acknowledgment}

This research was funded in part by the Indian government{'}s Visvesvaraya scheme for Ph.D.

\bibliographystyle{IEEEtran}
\bibliography{xyz.bib}

\begin{thebibliography}{10}
\providecommand{\url}[1]{#1}
\csname url@samestyle\endcsname
\providecommand{\newblock}{\relax}
\providecommand{\bibinfo}[2]{#2}
\providecommand{\BIBentrySTDinterwordspacing}{\spaceskip=0pt\relax}
\providecommand{\BIBentryALTinterwordstretchfactor}{4}
\providecommand{\BIBentryALTinterwordspacing}{\spaceskip=\fontdimen2\font plus
\BIBentryALTinterwordstretchfactor\fontdimen3\font minus
  \fontdimen4\font\relax}
\providecommand{\BIBforeignlanguage}[2]{{%
\expandafter\ifx\csname l@#1\endcsname\relax
\typeout{** WARNING: IEEEtran.bst: No hyphenation pattern has been}%
\typeout{** loaded for the language `#1'. Using the pattern for}%
\typeout{** the default language instead.}%
\else
\language=\csname l@#1\endcsname
\fi
#2}}
\providecommand{\BIBdecl}{\relax}
\BIBdecl

\bibitem{Sun2021-ap}
C.~Sun, X.~Wang, M.~A. Auwalu, S.~Cheng, and W.~Hu,
  ``\BIBforeignlanguage{en}{Organic thin film transistors‐based
  biosensors},'' \emph{\BIBforeignlanguage{en}{EcoMat}}, no. eom2.12094, 2021.

\bibitem{jain2020performance}
S.~Jain, A.~Joshi, and D.~Bharti, ``Performance investigation of organic thin
  film transistor on varying thickness of semiconductor material: An
  experimentally verified simulation study,'' \emph{Semiconductors}, vol.~54,
  no.~11, pp. 1483--1489, 2020.

\bibitem{ahmad2022novel}
R.~Ahmad, A.~JOSHI, and D.~Boolchandani, ``A novel instrumentation amplifier
  with high tunable gain and cmrr for biomedical applications,'' \emph{Turkish
  Journal of Electrical Engineering \& Computer Sciences}, vol.~30, no.~3, pp.
  996--1015, 2022.

\bibitem{Seo2020-08-ir}
G.~Seo, G.~Lee, M.~J. Kim, S.~H. Baek, M.~Choi, K.~B. Ku, C.~S. Lee, and S.~J.
  Kim, ``Correction to rapid detection of {COVID-19} causative virus
  ({SARS-CoV-2}) in human nasopharyngeal swab specimens using {FieldEffect}
  {Transistor-Based} biosensor,'' \emph{ACS Nano}, vol.~14, no.~9, pp.
  12\,257--12\,258,, 2020-08.

\bibitem{Saha2021-ep}
R.~Saha, Y.~Hirpara, and S.~Hoque, ``Sensitivity analysis on dielectric
  modulated ge-source {DMDG} {TFET} based label-free biosensor,'' \emph{IEEE
  Trans. Nanotechnol.}, vol.~20, pp. 552--560, 2021.

\bibitem{Dwivedi2020-ia}
P.~Dwivedi and R.~Singh, ``Investigation the impact of the gate work-function
  and biases on the sensing metrics of {TFET} based biosensors,'' \emph{Eng.
  Res. Express}, vol.~2, no.~2, p. 025043, 2020.

\bibitem{Hafiz1905-sg}
S.~A. Hafiz, I.~M. Ehteshamuddin, and S.~A. Loan, ``Dielectrically modulated
  {Source-Engineered} {Charge-Plasma-Based} {SchottkyFET} as a {Label-Free}
  biosensor,'' \emph{IEEE Transactions on Electron Devices}, vol.~66, no.~4,
  1905.

\bibitem{Mahalaxmi2020-12-01-kt}
B.~A. Mahalaxmi and G.~P. Mishra, ``Design and analysis of {Dual-Metal-Gate}
  {Double-Cavity} {Charge-Plasma-TFET} as a {Label-Free} biosensor,''
  \emph{IEEE Sensors Journal}, vol.~20, no.~23, pp. 13\,969--13\,975,,
  2020-12-01.

\bibitem{Dixit2021-yd}
A.~Dixit, D.~P. Samajdar, and N.~Bagga, ``Dielectric modulated finfet as a
  label-free biosensor: device proposal and investigation,'' \emph{Semicond.
  Sci. Technol.}, vol.~36, no.~9, p. 095033, 2021.

\bibitem{Wadhwa2019-bg}
G.~Wadhwa and B.~Raj, ``Design, simulation and performance analysis of {JLTFET}
  biosensor for high sensitivity,'' \emph{IEEE Trans. Nanotechnol.}, vol.~18,
  pp. 567--574, 2019.

\bibitem{Goel2021-lj}
A.~Goel, S.~Rewari, S.~Verma, S.~S. Deswal, and R.~S. Gupta, ``Dielectric
  modulated junctionless biotube {FET} ({DM-JL-BT-FET}) {Bio-Sensor},''
  \emph{IEEE Sens. J.}, vol.~21, no.~15, pp. 16\,731--16\,743, 2021.

\bibitem{Priyadarshani2021-tj}
K.~N. Priyadarshani and S.~Singh, ``\BIBforeignlanguage{en}{Ultra sensitive
  label-free detection of biomolecules using vertically extended drain double
  gate {Si0.5Ge0.5} source tunnel {FET}},'' \emph{\BIBforeignlanguage{en}{IEEE
  Trans. Nanobioscience}}, vol.~PP, no.~4, pp. 480--487, 2021.

\bibitem{Wang2019-bj}
N.~Wang, A.~Yang, Y.~Fu, Y.~Li, and F.~Yan,
  ``\BIBforeignlanguage{en}{Functionalized organic thin film transistors for
  biosensing},'' \emph{\BIBforeignlanguage{en}{Acc. Chem. Res.}}, vol.~52,
  no.~2, pp. 277--287, 2019.

\bibitem{Narang2015-ho}
R.~Narang, M.~Saxena, and M.~Gupta, ``Comparative analysis of
  dielectric-modulated {FET} and {TFET-based} biosensor,'' \emph{IEEE Trans.
  Nanotechnol.}, vol.~14, no.~3, pp. 427--435, 2015.

\bibitem{Kanungo2017-gd}
S.~Kanungo, S.~Chattopadhyay, K.~Sinha, P.~S. Gupta, and H.~Rahaman, ``A device
  simulation-based investigation on dielectrically modulated fringing
  field-effect transistor for biosensing applications,'' \emph{IEEE Sens. J.},
  vol.~17, no.~5, pp. 1399--1406, 2017.

\bibitem{Kumar2020-up}
A.~Kumar, M.~Roy, N.~Gupta, M.~M. Tripathi, and R.~Chaujar,
  ``\BIBforeignlanguage{en}{Dielectric modulated transparent gate thin film
  transistor for biosensing applications},''
  \emph{\BIBforeignlanguage{en}{Mater. Today}}, vol.~28, pp. 141--145, 2020.

\bibitem{Seo2020-vl}
G.~Seo, G.~Lee, M.~J. Kim, S.-H. Baek, M.~Choi, K.~B. Ku, C.-S. Lee, S.~Jun,
  D.~Park, H.~G. Kim, S.-J. Kim, J.-O. Lee, B.~T. Kim, E.~C. Park, and S.~I.
  Kim, ``\BIBforeignlanguage{en}{Rapid detection of {COVID-19} causative virus
  ({SARS-CoV-2}) in human nasopharyngeal swab specimens using field-effect
  transistor-based biosensor},'' \emph{\BIBforeignlanguage{en}{ACS Nano}},
  vol.~14, no.~4, pp. 5135--5142, 2020.

\bibitem{Zaki_undated-dx}
T.~Zaki, S.~Scheinert, I.~H{\"o}rselmann, R.~R{\"o}del, F.~Letzkus, H.~Richter,
  U.~Zschieschang, H.~Klauk, and J.~N. Burghartz, ``Accurate capacitance
  modeling and characterization of organic thin-film transistors``,''
  \emph{IEEETrans. Electr. Dev}, vol.~61, no. 1,2014.

\bibitem{Singh2020-04-up}
A.~K. Singh, A.~Pandey, and P.~Chakrabarti,
  ``Poly[2,5-bis(3tetradecylthiophen-2-yl) thieno [3,2-b] thiophene] organic
  polymer {Based-Interdigitated} channel enabled thin film transistor for
  detection of selective low ppm ammonia sensing at 25°c,'' \emph{IEEE Sensors
  Journal}, vol.~20, no.~8, pp. 4047--4055,, 2020-04.

\bibitem{Rashid2021-08-15-fk}
S.~Rashid, F.~Bashir, and F.~A. Khanday, ``Dielectrically modulated
  {Label-Free} metal controlled organic thin film transistor for biosensing
  applications,'' \emph{IEEE Sensors Journal}, vol.~21, no.~16, pp.
  18\,318--18\,325,, 2021-08-15.

\bibitem{Bartic2003-ev}
C.~Bartic, A.~Campitelli, and S.~Borghs, ``Field-effect detection of chemical
  species with hybrid organic/inorganic transistors,'' \emph{Appl. Phys.
  Lett.}, vol.~82, no.~3, pp. 475--477, 2003.

\bibitem{Khan2011-iq}
H.~U. Khan, J.~Jang, J.-J. Kim, and W.~Knoll, ``\BIBforeignlanguage{en}{Effect
  of passivation on the sensitivity and stability of pentacene transistor
  sensors in aqueous media},'' \emph{\BIBforeignlanguage{en}{Biosens.
  Bioelectron.}}, vol.~26, no.~10, pp. 4217--4221, 2011.

\bibitem{Minamiki2017-iw}
T.~Minamiki, Y.~Sasaki, S.~Tokito, and T.~Minami, ``Label-free direct
  electrical detection of a histidine-rich protein with subfemtomolar
  sensitivity using an organic field-effect transistor,'' \emph{ChemistryOpen},
  vol.~6, pp. 472--475, 2017.

\bibitem{Loi2005-pk}
A.~Loi, I.~Manunza, and A.~Bonfiglio, ``Flexible, organic, ion-sensitive
  field-effect transistor,'' \emph{Appl. Phys. Lett.}, vol.~86, no.~10, p.
  103512, 2005.

\bibitem{Yan2009-oc}
F.~Yan, S.~M. Mok, J.~Yu, H.~L.~W. Chan, and M.~Yang,
  ``\BIBforeignlanguage{en}{Label-free {DNA} sensor based on organic thin film
  transistors},'' \emph{\BIBforeignlanguage{en}{Biosens. Bioelectron.}},
  vol.~24, no.~5, pp. 1241--1245, 2009.

\bibitem{joshi2022label}
S.~Jain and A.~Joshi, ``Label-free detection of biomolecules using
  dielectric-modulated top contact bi-layer electrodes organic thin film
  transistor,'' 2022.

\bibitem{Syu2018-06-gd}
Y.~C. Syu, W.~E. Hsu, and C.~T. Lin, ``Field-effect transistor biosensing:
  Devices and clinical applications,'' \emph{ECS Journal of Solid State Science
  and Technology}, vol.~7, no.~7, pp. 3196 3207,, 2018-06.

\bibitem{Zan2012-zf}
H.-W. Zan, W.-W. Tsai, Y.-R. Lo, Y.-M. Wu, and Y.-S. Yang, ``Pentacene-based
  organic thin film transistors for ammonia sensing,'' \emph{IEEE Sens. J.},
  vol.~12, no.~3, pp. 594--601, 2012.

\bibitem{sadighbayan2021laser}
D.~Sadighbayan, A.~Minhas-Khan, and E.~Ghafar-Zadeh, ``Laser-induced
  graphene-functionalized field-effect transistor-based biosensing: A potent
  candidate for covid-19 detection,'' \emph{IEEE Transactions on
  NanoBioscience}, vol.~21, no.~2, pp. 232--245, 2021.

\bibitem{ahmad2021design}
R.~Ahmad, A.~M. Joshi, D.~Boolchandani, and T.~Varma, ``Design of potentiostat
  and current mode read-out amplifier for glucose sensing,'' in \emph{2021 IEEE
  International Symposium on Smart Electronic Systems (iSES)(Formerly
  iNiS)}.\hskip 1em plus 0.5em minus 0.4em\relax IEEE, 2021, pp. 64--69.

\bibitem{gupta2020surface}
M.~Gupta, S.~Santermans, G.~Hellings, L.~Lagae, K.~Martens, and W.~Van~Roy,
  ``Surface charge modulation and reduction of non-linear electrolytic
  screening in fet-based biosensing,'' \emph{IEEE Sensors Journal}, vol.~21,
  no.~4, pp. 4143--4151, 2020.

\bibitem{Wadhwa2018-05-uk}
G.~Wadhwa and B.~Raj, ``Label-free detection of biomolecules using
  charge-plasma-based gate underlap dielectric modulated junctionless {TFET},''
  \emph{J. Electron. Mater}, vol.~47, no.~8, pp. 4883--4893,, 2018-05.

\bibitem{Hafiz2019-02-dd}
S.~A. Hafiz, I.~M. Ehteshamuddin, and S.~A. Loan, ``Dielectrically modulated
  {Source-Engineered} {Charge-Plasma-Based} {SchottkyFET} as a {Label-Free}
  biosensor,'' \emph{IEEE Transactions on Electron Devices}, vol.~66, no.~4,
  pp. 1905--1910,, 2019-02.

\bibitem{Venkatesh2017-09-vs}
P.~Venkatesh, K.~Nigam, S.~Pandey, D.~Sharma, and P.~N. Kondekar, ``A
  dielectrically modulated electrically doped tunnel {FET} for application of
  label-free biosensor,'' \emph{Superlattices and Microstructures}, vol. 109,
  pp. 470--479,, 2017-09.

\bibitem{Singh2017-lg}
D.~Singh, S.~Pandey, K.~Nigam, D.~Sharma, D.~S. Yadav, and P.~Kondekar, ``A
  charge-plasma-based dielectric-modulated junctionless {TFET} for biosensor
  label-free detection,'' \emph{IEEE Trans. Electron Devices}, vol.~64, no.~1,
  pp. 271--278, 2017.

\bibitem{Kraft2016-we}
U.~Kraft, K.~Takimiya, M.~J. Kang, R.~R{\"o}del, F.~Letzkus, J.~N. Burghartz,
  E.~Weber, and H.~Klauk, ``Detailed analysis and contact properties of
  low-voltage organic thin-film transistors based on dntt and its didecyl and
  diphenyl derivatives{'}{'},'' \emph{Organic Electronics}, vol.~35, pp.
  33--40,, 2016.

\bibitem{Im2007-ha}
H.~Im, X.-J. Huang, B.~Gu, and Y.-K. Choi, ``\BIBforeignlanguage{en}{A
  dielectric-modulated field-effect transistor for biosensing},''
  \emph{\BIBforeignlanguage{en}{Nat. Nanotechnol.}}, vol.~2, no.~7, pp.
  430--434, 2007.

\bibitem{Wadhwa2018-yy}
G.~Wadhwa and B.~Raj, ``\BIBforeignlanguage{en}{Label free detection of
  biomolecules using charge-plasma-based gate underlap dielectric modulated
  junctionless {TFET}},'' \emph{\BIBforeignlanguage{en}{J. Electron. Mater.}},
  vol.~47, no.~8, pp. 4683--4693, 2018.

\end{thebibliography}

\end{document}